# A computationally-efficient construction for the matrix-based key distribution in sensor network

Abedelaziz Mohaisen
Electronics and Telecommunication Research Institute
Daejeon 305-700, Korea
a.mohaisen@etri.re.kr

## Abstract

*This paper introduces a variant for the symmetric matrix-based key distribution in sensor network introduced by Du et al. Our slight modification shows that the usage of specific structures for the public matrix instead of fully random matrix with elements in $\mathbb{Z}_q$ can reduce the computation overhead for generating the public key information and the key itself. An intensive analysis followed by modified scheme demonstrates the value of our contribution in relation with the current work and show the equivalence of the security*

## 1. Introduction

Key pre-distribution (KPD) is a challenging issue in deploying the symmetric key cryptography for wireless sensor network (WSN). In the KPD, a set of keys or keying material is assigned to each node to ensure a secure communication between the nodes in the real time manner. Due to the absence of the infrastructure, traditional KPD methods such like the key distribution center (KDC), where a centralized authority for key distribution exists, are discarded and considered infeasible.

To solve the above problem, several works and schemes have been introduced. These schemes range from the graph-based cryptographic keys assignment such like the works in [4, 5, 2] to the more sophisticated online key generation schemes such like works in [3, 6, 8, 7]. In this paper, we review of these schemes and provide a construction on it to reduce its usage of resources while maintaining the same level of security. The revisited scheme is DDHV in introduced by Du et al. in [3]

We mainly introduce a construction based on the DDHV scheme to reduce the used computation overhead with a small additional communication overhead. Our main contributions are summarized as follows: (1) We introduce a special construction that reduces the computation overhead with a small additional communication overhead. (2) We show a concrete evaluation for the soundness of the scheme, the security achieved and the resources evaluation. (3) To show a comparison between the modified DDHVscheme and the original work.

The rest of this paper is organized as follows: section 2 introduces an overview of DDHV scheme followed by our scheme in section 3, section 4 introduces the analysis of both schemes where we show the overhead evaluation in terms of communication, computation and memory followed by the security analysis. Finally, section 5 draws a concluding remarks.

## 2. DDHV scheme

The DDHV scheme in [3] utilizes Blom's scheme in [1] with Eschenauer and Gligor's random key assignment concept in [4]. Roughly speaking, both DDHV and Blom's schemes are based on the symmetry concept of matrices to provide symmetric pairwise keys for the pairs of communicating nodes. DDHV scheme differs in that it utilizes multiple spaces for generating the key. In this paper, we will explain the discuss the symmetric matrix-based component of DDHVas our modification is only related to that part.

Naively, a symmetric matrix of size $N \times N$ can be used for storing the different $N^2$ keys used for securing communication within the entire network of size $N$ where each node $s_i$ can have a row in that matrix. If two nodes $s_i$ and $s_j$ would like to communicate securely, they use the corresponding elements for encrypting and decrypting the communication traffic symmetrically. That is, $E_{ij}$ is used in $s_i$'s side and $E_{ji}$ is used in $s_j$'s side where both are equal according to the symmetry of the main matrix. To reduce the memory requirements, a linear algebraic-based construction is introduced where the size of the square matrix is reduced into $\lambda \ll N$. In Blom (and therefore in DDHV) scheme [1], the following are defined: a public ma-trix $\mathbf{G}$ of size $(\lambda+1) \times N$ and a private symmetric matrix $\mathbf{D}$ of size $(\lambda+1) \times (\lambda+1)$ where elements of $\mathbf{G}$ and $\mathbf{D}$ are randomly generated in the finite field $\mathbb{Z}_q$. Also, a matrix $\mathbf{A}$ is defined and computed as $\mathbf{A} = (\mathbf{DG})^T$ which is of size $N \times (\lambda+1)$. For any node $s_i$, the corresponding row $\mathbf{A}_r(i)$ from $\mathbf{A}$ and the corresponding column $\mathbf{G}_c(i)$ from $\mathbf{G}$ are selected and loaded in the node's memory. When $s_i$ and $s_j$ need to communicate securely, they exchange $\mathbf{G}_c(i)$ and $\mathbf{G}_c(j)$ respectively and then $k_{ij} = \mathbf{A}_r(i) \times \mathbf{G}_c(j)$ is computed in the side of $s_i$ and $k_{ji} = \mathbf{A}_r(j) \times \mathbf{G}_c(i)$ is computed in the side of $s_j$. Obviously, the resulting keys are equal due to the symmetry property of the matrix $\mathbf{D}$.

To reduce the communication overhead, the DDHV scheme introduced the a construction of $\mathbf{G}$ based on Vandermonde matrix which can be represented as in (1) where each node stores the corresponding field element in the matrix and generate the whole column from that value. Obviously, to construct corresponding column from the given value, $\lambda$ number of multiplications over $\mathbb{Z}_q$ are required. Similarly, to generate the key by multiplying $\mathbf{A}_r$ by $\mathbf{G}_c$, another $\lambda$ number of multiplications over $\mathbb{Z}_q$ is required.

$$\mathbf{G} = \begin{bmatrix} 1 & 1 & 1 & \ldots & 1 \\ s & (s^2) & (s^3) & \ldots & (s^N) \\ s^2 & (s^2)^2 & (s^3)^2 & \ldots & (s^N)^2 \\ \vdots & \vdots & \vdots & \ddots & \vdots \\ s^\lambda & (s^2)^\lambda & (s^3)^\lambda & \ldots & (s^N)^\lambda \end{bmatrix} \quad (1)$$

## 3. Modified Scheme (OR-DDHV)

Our modification for the above DDHV scheme relies in reducing the computation overhead with a slight increment in the used communication overhead while maintaining the same security level. That is, we re-design the public matrix $\mathbf{G}$ in such a way that maximize the number of zeros leading to that the inner multiplications used for generating the key are made as few as possible. Also, when several elements are set to zero in the matrix $\mathbf{G}$, additional overhead required for reconstructing the public information when exchanging it will be discarded.

Let the matrix (2) represents $\mathbf{G}^T$ in which each row has only two nonzero values. According to the above DDHV scheme, each node has a column in $\mathbf{G}$ represented by two non-zero values. Based on the $\mathbf{G}^T$, we define the ofline and online phases in the following sections.

$$\mathbf{G}^T = \begin{bmatrix} g_{11} & g_{12} & 0 & \ldots & 0 & 0 \\ 0 & g_{22} & g_{23} & \ldots & 0 & 0 \\ \vdots & \vdots & \vdots & \ddots & \vdots & \vdots \\ 0 & 0 & 0 & \ldots & g_{(\lambda-1)(\lambda-1)} & g_{(\lambda-1)(\lambda)} \\ 0 & g_{(\lambda)(\lambda+1)} & 0 & \ldots & 0 & g_{(\lambda)(\lambda)} \\ \vdots & \vdots & \vdots & \ddots & \vdots & \vdots \end{bmatrix} \quad (2)$$

### 3.1. Offline Phase

1. The administrator generates a symmetric matrix $\mathbf{D}$ of size $\lambda \times \lambda$ with elements in $\mathbb{Z}_q$ and the public matrix $\mathbf{G}$ of size $\lambda \times N$ with elements in $\mathbb{Z}_q$ where $\mathbf{G}$ satisfies the above restrictions.

2. The administrator computes $\mathbf{A} = \mathbf{G}^T\mathbf{D}$. The resulting $\mathbf{A}$ is of size $N \times \lambda$ and therefore its elements are in $\mathbb{Z}_q$.

3. For each node $s_i$, the administrator assigns the row with index $i$ from the matrix $\mathbf{A}$ (e.g., $\mathbf{A}_r(i)$) and column with index $i$ from the matrix $\mathbf{G}$ (i.e., $\mathbf{G}_c(i)$).

### 3.2. Online phase

The online phase consists of the following steps:

1. Firstly, two nodes $s_i$ and $s_j$ exchange their public columns $\mathbf{G}_c(i)$ and $\mathbf{G}_c(i)$ which can be represented as two non-zero values in $\mathbb{Z}_q$ and denoted as $g_{1i}, g_{2i}, g_{1i}, g_{2j}$.

2. In a vector $\mathbf{G}_c(j)$ with zero elements, the node $s_i$ sets the received $g_{1i}$ and $g_{2j}$ from the node $s_j$ with the identifier $j$ into following positions in $\mathbf{G}_c(j)$:

$$\mathbf{G}_c(j)[j \mod \lambda] \leftarrow g_{1j}.$$
$$\mathbf{G}_c(j)[(j+1) \mod \lambda] \leftarrow g_{2j}.$$

3. Similarly, the node $s_j$ reconstruct $\mathbf{G}_c(i)$ by plugging the received values $g_{1i}, g_{2i}$ in the following positions:

$$\mathbf{G}_c(i)[i \mod \lambda] \leftarrow g_{1i}$$
$$\mathbf{G}_c(i)[(i+1) \mod \lambda] \leftarrow g_{2i}$$

4. The node $s_i$ computes $k_{ij} = \mathbf{A}_r(i)\mathbf{G}_c(j)$.

5. The node $s_j$ computes $k_{ji} = \mathbf{A}_r(j)\mathbf{G}_c(i)$.

## 4. Analysis

### 4.1. Limitations on the Network Size

The maximum supported network size in our scheme is merely dependent on the parameters $N$ and $\lambda$. In order to avoid a possible collision and maintain the vectors of $\mathbf{G}$ independent, maximum network size is set to $N = 2 \times \lambda$.

## 4.2. Equivalence of keys

We can simply show that the generated key are equal. That is equivalent to showing that if $\mathbf{D}$ symmetric then $\mathbf{B} = \mathbf{G}^T\mathbf{D}\mathbf{G}$ is also symmetric and therefore the resulting keys are equal at both sides of $s_i$ and $s_j$. To show the symmetry of $\mathbf{B}$, it is enough to demonstrate that $\mathbf{B} = \mathbf{B}^T$. That is, $\mathbf{B}^T = (\mathbf{G}^T\mathbf{AG})^T = \mathbf{G}^T(\mathbf{G}^T\mathbf{A})^T = \mathbf{G}^T\mathbf{A}^T\mathbf{G} = \mathbf{G}^T\mathbf{A}^T\mathbf{G} = \mathbf{B}$. Since both $k_{ij}$ and $k_{ji}$ are elements in $\mathbf{B}$ which is symmetric, both keys are equal.

Let $a_{ij}, d_{ij}$ and $g_{ij}$ be the $(i, j)$ elements in the matrices $\mathbf{A}, \mathbf{D}$ and $\mathbf{G}$ respectively. Also, let $\mathbf{A} = (\mathbf{DG})^T$. From which we would like to show that $k_{ij} = A_r(i)G_c(j)$ and $k_{ji} = A_r(j)G_c(i)$ are equal.

*Proof.*

We can write $a_{ij}$ with corresponding to its multipliers as follows: $a_{ij} = (\sum_{k=1}^{\lambda} d_{ik}g_{ki})^T = (\sum_{k=1}^{\lambda} d_{ik}g_{ki})$ From which we can write $A_r(i) = [a_{1i}, a_{2i}, \ldots] = \left[\sum_{k=1}^{\lambda} d_{1k}g_{ki}, \sum_{k=1}^{\lambda} d_{2k}g_{ki}, \ldots\right]$ and $G_c(j) = [(g_{1j}, g_{2j}, \ldots)]$. The resulting of $A_r(i) \times G_c(j)$ can be written as follows:

$$A_r(i)G_c(j) = \sum_{l=1}^{\lambda}\left(\sum_{k=1}^{\lambda} d_{lk}g_{ki}\right)g_{lj} \quad (3)$$

Similarly, we can show that $A_r(j)G_c(i) = \sum_{l=1}^{\lambda}(\sum_{k=1}^{\lambda} d_{lk}g_{kj})g_{li}$. Now, we would like to check whether $A_r(j)G_c(i) = A_r(i)G_c(j)$ for any $i \neq j$. That is, we would like to show the following equality.

$$\sum_{l=1}^{\lambda}\left(\sum_{k=1}^{\lambda} d_{lk}g_{ki}\right)g_{lj} \stackrel{?}{=} \sum_{l=1}^{\lambda}\left(\sum_{k=1}^{\lambda} d_{lk}g_{kj}\right)g_{li} \quad (4)$$

By Taking the right side in (4) and change the index of the summations we get the that: $\sum_{l=1}^{\lambda}(\sum_{k=1}^{\lambda} d_{lk}g_{kj})g_{li} = \sum_{k=1}^{\lambda}(\sum_{l=1}^{\lambda} d_{kl}g_{li})g_{ki} = \sum_{k=1}^{\lambda}(\sum_{l=1}^{\lambda} d_{lk}g_{li})g_{ki}$.

Because $\mathbf{D}$ is symmetric, $g_{li} = g_{il}$, therefore the above can be rewritten as: $\sum_{l=1}^{\lambda} d_{l1}g_{lj}g_{1i} + \sum_{l=1}^{\lambda} d_{l2}g_{lj}g_{2i} + \cdots = (d_{11}g_{1j}g_{1i} + d_{21}g_{2j}g_{1i} + d_{31}g_{3j}g_{1i} + \ldots) + (d_{12}g_{1j}g_{2i} + d_{22}g_{2j}g_{2i} + d_{32}g_{3j}g_{2i} + \ldots) + (d_{13}g_{1j}g_{3i} + d_{23}g_{2j}g_{3i} + d_{33}g_{3j}g_{3i} + \ldots) + \ldots$. By resuming and arranging the terms we get the following:

$$= g_{1j}\sum_{k=1}^{\lambda} d_{1k}g_{ki} + g_{2j}\sum_{k=1}^{\lambda} d_{2k}g_{ki} + \ldots$$

$$= \sum_{l=1}^{\lambda} g_{lj}\sum_{k=1}^{\lambda} d_{lk}g_{ki} = \sum_{l=1}^{\lambda}\left(\sum_{k=1}^{\lambda} d_{lk}g_{ki}\right)g_{lj} \quad (5)$$

From (3) and (5), we get that (4) holds. □

## 4.3. Resources overhead

- **Communication overhead:** The communication in the OR-DDHV scheme is $2\log_2 2^q = 2 \times q$ while it is $q$ bits in the DDHV scheme when transferring a single field value from which the corresponding column in $\mathbf{A}$ is generated.

- **Computation overhead:** The computation overhead in DDHV and OR-DDHV is two parts. First the first part is required for reconstructing the public information from the field element and the second part is required for computing the inner product to generate the symmetric key.

  - **Column's reconstruction computation:** The computation required in OR-DDHV scheme to reconstruct the corresponding column is negligible while it is $\lambda$ number of multiplications in the field $\mathbb{Z}_q$ in DDHV scheme. That is, when $\lambda$ is large, the number of computations over $q$ will be also large. To illustrate how the reconstruction works for the case of DDHV scheme, given $s^i$, any element in the column is the result of multiplying the two previous elements. That is, $s^i = 1 \times s^i$, $(s^i)^2 = s^i \times s^i$ and so on.

  - **Computation for inner product:** The computation for the inner product between the column from $\mathbf{G}$ and the row from $\mathbf{A}$ to obtain the symmetric key is 2 multiplications in our scheme since only two values are non-zero in $\mathbf{G}$'s corresponding column. On contrast, $\lambda$ number of multiplications in the field $\mathbb{Z}_q$ are required in the case of DDHV scheme.

  To sum up, the required computation overhead in term of multiplications in $\mathbb{Z}_q$ is 2 multiplications for OR-DDHV and $2\lambda$ multiplications for DDHV.

- **Memory overhead:** For simplicity, we consider that the required memory is only for storing the corresponding row in $\mathbf{A}$ for the node $s_i$ in its memory. Recalling that the elements of $\mathbf{A}$ are in $\mathbb{Z}_q$ and the length of each row in $\mathbf{A}$ is $\lambda$ elements, the required memory in OR-DDHV is same like the required memory in DDHV which is equal to $\lambda \times q$ bit.

A summary of the comparison in terms of the required resources is shown in Table 1. Note that though the communication overhead in OR-DDHV is higher than in DDHV, it is still constant since $q$ is fixed to accumulate the proper length of key. On contrast, the computation in the OR-DDHV is constant while it increase linearly according to the security parameter $\lambda$ in DDHV.

**Table 1. Comparison between** DDHV **and** OR-DDHV **in term of the used resources where communication and memory are in bit per node and computation is in term of multiplications in the finite field** $\mathbb{Z}_q$**.**

| algorithm | communication | computation | memory |
|---|---|---|---|
| DDHV | $q$ | $2\lambda$ | $\lambda \times q$ |
| OR-DDHV | $2q$ | $2$ | $\lambda \times q$ |

### 4.4 Security Analysis

The security analysis follows the analysis shown in DDHV or Blom work. That is, the system is $\lambda$-secure which leads to that an adversary needs to know $\lambda$ number of different linearly independent elements (i.e., rows or columns) from the key generation construction to be able to know the keys between uncompromised nodes. Recall $\mathbf{G}$ in (2), $\mathbf{A}$, and $\mathbf{D}$ defined above. Also recall that $a_{ij}$ and $d_{ij}$ are the $(i,j)$ elements of $\mathbf{A}$ and $\mathbf{D}$ respectively. Now we can define $\mathbf{A}_r(i)$ as $\mathbf{A}_r(i) = \begin{bmatrix} a_{i1} & a_{i2} & \ldots & a_{i\lambda} \end{bmatrix}$ where $a_{ij} = \left(\sum_{k=1}^{\lambda} d_{ik} g_{ki}\right)^T = \left(\sum_{k=1}^{\lambda} d_{ik} g_{ki}\right)$. The above $\mathbf{A}$ can be rewritten as:

$$\mathbf{A} = \begin{bmatrix} (g_{11}d_{11} + g_{12}d_{21}) & (g_{11}d_{12} + g_{12}d_{22}) & \ldots \\ (g_{22}d_{21} + g_{23}d_{31}) & (g_{22}d_{22} + g_{23}d_{32}) & \ldots \\ \vdots & \vdots & \ddots \end{bmatrix} \quad (6)$$

An adversary who would like to attack the above system must first reconstruct the proper $\mathbf{D}$. Since $\mathbf{D}$ is in $\mathbb{Z}^{\lambda \times \lambda}$, $\lambda^2$ number of linear equations are required for reconstructing it. That is, given that $\mathbf{G}$, the systematic structure of $\mathbf{A}$ and $\mathbf{G}$, and the symmetric property of $\mathbf{D}$ is publicly known information to the adversary, the adversary can obtain $\lambda$ different linear equations by attacking a single node and reconstructing the different equations representing the row $\mathbf{A}_r(i)$. By attacking the nodes with the ID 1, the attacker will have the following: $a_{11} = g_{11}d_{11} + g_{12}d_{21}$, $a_{12} = g_{11}d_{12} + g_{12}d_{22}$, $a_{13} = g_{11}d_{13} + g_{12}d_{23}, \ldots$.

By repeating the physical attack to $\lambda$ different nodes, the adversary can construct $\lambda^2$ linear equation with $\lambda^2$ variables that can be solved to recover the whole private matrix $\mathbf{D}$ and construct any pairwise key between any pair of uncompromised nodes by just observing their public information. Note that the existence of multiple zeros in the $\mathbf{G}$ will not reduce the hardness of solving the above linear system since the different elements of the matrix $\mathbf{D}$ always exist in the resulting linear construction in $\mathbf{A}$. In DDHV scheme, however, all variables (represent by the different $d$'s) appear in each equation rather than the two variables in each as shown above.

## 5 Conclusion

This paper introduced a variant for DDHV work. We demonstrated that the usage of the orthogonal matrix instead of fully random matrix with elements in $\mathbb{Z}_q$ will lead to a great reduction in the overhead represented by the computation required for generating the public key material and the key itself.